\documentclass[twocolumn,showpacs,preprintnumbers,amsmath,amssymb]{revtex4}

\usepackage{graphicx}
\usepackage{bm}

\begin{document}

\title{ Magnetic-dipole transitions in highly-charged ions as a basis of ultra-precise optical clocks }

\author{V. I. Yudin\footnote{E-mail address: viyudin@mail.ru}, A. V. Taichenachev}
\affiliation{Institute of Laser Physics SB RAS, pr. Akademika Lavrent'eva 13/3, Novosibirsk, 630090, Russia}
\affiliation{Novosibirsk State University, ul. Pirogova 2, Novosibirsk, 630090, Russia}
\affiliation{Novosibirsk State Technical University, pr. Karla Marksa 20, Novosibirsk, 630073, Russia}
\affiliation{Russian Quantum Center, Skolkovo, Moscow Reg., 143025, Russia}
\author{A. Derevianko}
\affiliation{Department of Physics, University of Nevada, Reno,
Nevada 89557, USA}

\date{\today}

\begin{abstract}
We evaluate the feasibility of using magnetic-dipole (M1) transitions in highly-charged ions as a basis of an optical atomic clockwork of exceptional accuracy.
We consider a range of possibilities, including M1 transitions between clock levels of the same fine-structure and hyperfine-structure manifolds.
In highly charged ions these transitions lie in the optical part of the spectra and can be probed with lasers. The most direct advantage of our proposal comes from the low degeneracy of clock levels and the simplicity of atomic structure in combination with negligible quadrupolar shift.
We demonstrate that such clocks can have projected fractional accuracies below the $10^{-20}-10^{-21}$ level for all common systematic effects, such as black-body radiation, Zeeman, AC-Stark and quadrupolar shifts. Notice that  usually-employed hyperfine clock transitions lie in the microwave spectral region. Our proposal moves such transitions to the optical domain. As the hyperfine transition frequencies depend on the fine-structure constant, electron-to-proton mass ratio, and nuclear magnetic moment, our proposal expands the range of experimental schemes for probing  space and time variations  of fundamental constants.
\end{abstract}


\maketitle

Development of  optical atomic clocks has seen truly impressive advances over the past decade.
Two approaches have competed and  complemented each other: singly-charged trapped ions~\cite{Chou_2010} and an ensemble of neutral atoms trapped in designer ``magic''  optical lattices~\cite{Katori1}. These experiments rapidly expanded the frontiers of accuracy and stability.
 Presently  ion clocks ($^{23}$Al$^+$ \cite{Chou_2010}, $^{171}$Yb$^+$
\cite{Huntemann_2012} and $^{88}$Sr$^+$ \cite{Madej_2012})  have reached the $10^{-17}$ fractional stability while lattice clocks have demonstrated the stability at the $10^{-18}$ level (Yb~\cite{Hinkley_2013}, Sr~\cite{BloNicWil14}).
The projected accuracy of these clocks is at the level of 10$^{-17}$--10$^{-18}$.  Such accuracies are anticipated to have implications both for practical (e.g.,  relativistic geodesy) and fundamental applications (testing time and space variation of fundamental constants~\cite{Rosenband} and search for topological dark matter~\cite{DerPos14}).

For any atomic clockwork one has to carefully minimize environmental perturbations of  a quantum oscillator.
For example, clock frequencies can be  affected by the thermal bath of  black-body radiation (BBR), electric-quadrupole couplings to the trapping and residual electromagnetic fields, ambient magnetic fields and through various Stark shifts. Novel classes of clocks: nuclear clocks \cite{Peik_2003} and clocks employing highly-charged ions \cite{Derevianko_2012} are naturally impervious to such perturbations. Nuclear clocks utilizing  the isomeric transition in $^{229}$Th nucleus are estimated \cite{Campbell_2012} to have the fractional accuracy at the 10$^{-19}$  level. Here the improvement comes due to the small size of a nucleus and thereby reduced couplings to environmental perturbation.
However, the low-energy $^{229}$Th nuclear transition has not been  observed yet and its frequency is estimated with a large uncertainty ($\sim$~eV). As an alternative to reaching accuracies similar to that of the nuclear clock, Ref.~\cite{Derevianko_2012} proposed using optical transitions in highly-charged ions (HCI).
These authors considered electric-quadrupole (E2) transitions between terms arising from the 4$f^{12}$ electronic configuration. As for the nucleus the key advantage of the HCIs is the reduced size of the quantum oscillator. Indeed, compared to a neutral atom, the HCI electronic cloud is shrunk by the factor of $Z_i$ ($Z_i$ is the residual charge of the ion, related to the nuclear charge $Z$ and the total number of electrons $N_e$ as $Z_i = Z-N_e$.) Moreover, gross energy intervals nominally grow as $Z_i^2$ with increasing $Z_i$. This leads to substantial suppression {of AC-Stark and BBR shifts as $1/Z_i^4$. The quadrupolar moments of electronic clouds scale as $1/Z^2_i$ reducing   quadrupolar shifts. Additional advantage of the 4$f^{12}$ manifold stems from the possibility~\cite{Derevianko_2012}  of choosing clock states that in addition to the mentioned $1/Z^2_i$ overall suppression minimize differential quadrupolar shift (i.e., accidental ``arithmetic'' suppression). Other possible HCI clock transitions were recently proposed in \cite{Safronova_2014}.

\begin{figure}[h]
\centerline{\scalebox{0.4}{\includegraphics{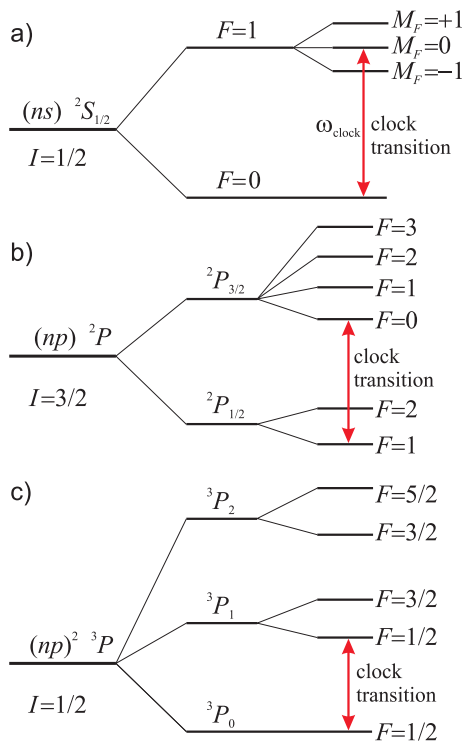}}}\caption{
Illustrative examples of magnetic-dipole clock transitions in mono-valent (panels (a) and (b)) and divalent (c) highly-charged ions.
The ordering of the hyperfine levels and values of hyperfine intervals depends on the signs and relative sizes of hyperfine structure constants.
 A particular choice of nuclear spins in panels (b) and (c) and hyperfine states forming  clock transitions  eliminates quadrupolar shifts.
\label{Fig1}}
\end{figure}

In this Letter we propose an alternative approach to the new generation of optical clocks based on highly-charged ions. In contrast to the HCI proposals \cite{Derevianko_2012,Safronova_2014} we employ the following strategy: we focus on the magnetic-dipole (M1) optical clock
transitions between states belonging either to the same fine- or hyperfine-structure manifolds attached to the ground electronic state. We argue that this choice substantially simplifies clock level structure removing degeneracies and enabling simpler clock initialization and readout. Also, substantial quadrupolar shifts can be either strongly suppressed or fully eliminated. We argue that our proposed HCI clock based on M1 optical transitions  can be considered as a suitable candidate for the next generation of  frequency standards with fractional uncertainty at the 10$^{-19}$ level.

For example,  hydrogen-like ($N_e$=1) and alkali-like  ($N_e$=3,11,29,...) HCIs
have a single valence electron outside a closed-shell core. Their ground electronic term is $^{2}S_{1/2}$ with the total angular momentum $J=1/2$.
Interaction of the nuclear magnetic moment with the electron spin splits the ground state into two hyperfine levels
characterized by the angular momentum $F=I\pm 1/2$, where $I\ne0$ is the nuclear spin (see Fig.\ref{Fig1}a).
M1 transition between these two levels for HCI with $Z_i>60$  can indeed lie in the optical spectral region (see Table~\ref{T1}) and thereby can be used as a basis of an optical clock. It is worth emphasizing that  while in neutral atoms hyperfine transitions are in the microwave domain, it is the scaling with $Z_i$ that pushes these frequencies into optical spectrum for the HCIs.

The number of candidate HCIs is vastly expanded when we further focus on M1 transitions connecting levels of fine-structure manifolds
of the ground electronic state. Fine structure emerges due to relativistic interactions  that are orders of magnitude more sizable than
couplings to nuclear spins. Thereby the fine-structure intervals are much larger than the hyperfine splittings and optical transition frequencies are attained for smaller values of residual charge $Z_i>5$. Some of possibilities are displayed in Fig.~\ref{Fig1}b,c, where we consider monovalent and divalent ions.  A particular choice of nuclear spins and hyperfine states forming  clock transitions fully eliminates quadrupolar shifts discussed later.

Our analysis demonstrates that all the common systematic effects for HCI M1 clock transitions
can be substantially suppressed below the  $10^{-20}$ level of fractional accuracy.
As an illustration we start by evaluating systematic clock shifts for the simplest example of hydrogen-like ions with the nuclear spin $I=1/2$.
The hyperfine structure of the ground state consists of two levels $F=1$ and $F=0$ (see  Fig.\ref{Fig1}a). The $F=0$ level lacks
Zeeman components eliminating the complexities of state preparation as otherwise required for degenerate levels (see, e.g.  \cite{Chou_2010}).

{\em Quadrupole shift in inhomogenous electric fields.} --
In ion clocks RF field trapping is essential for eliminating
Doppler broadening. The trapping fields are necessarily inhomogeneous. The gradients of the trapping electric field couple to the quadrupole moments (Q-moments) of clock states leading to undesirable shifts of clock frequencies. Our choice of the $J=1/2$ clock states warrants vanishing contribution of the electrons to the Q-moments due to the angular selection rules (quadrupole operator is a rank 2 irreducible tensor). Similarly for the  $I=1/2$ nuclei, the nuclear Q-moment vanishes for the same reasons. Even for the $I>1/2$ nuclei the quadrupole  coupling is exceptionally small as the nuclear  Q-moments (proportional to size squared) are many orders of magnitude ($\sim 10^{10}$) smaller than the electronic Q-moments.

{\em Zeeman shifts in magnetic field.} --
Application of magnetic field is required to lift the degeneracy of the $F=1$ clock states. However, the B-field shifts the clock transition.
The transition between the $M_F=0$ clock states in Fig.~\ref{Fig1}a  is clearly unaffected by the linear Zeeman shifts, $\Delta^{(1)}_{Z}$=0,
and we need to evaluate the second-order shift. Its value relative to the clock  frequency $\omega_\text{clock}$ reads
\begin{equation}\label{Zeeman}
    \frac{\Delta^{(2)}_{Z}}{\omega_\text{clock}}=\frac{2\mu^2_\text{B} |{\bf H}|^2}{\hbar^2\omega^2_\text{clock}} \, ,
\end{equation}
where $\mu^{}_\text{B}$ is the Bohr magneton and $H$ is the B-field value.
For order-of-magnitude estimates we fix the clock frequency at
$\omega_\text{clock}/2\pi$=$2\times$$10^{14}$~Hz (
$\lambda_\text{clock}\approx 1.5$~$\mu$m.) For a B-field  $\mu_\text{B} |{\bf H}|/ (2\pi\hbar) \sim10^4$~Hz the relative clock shift (\ref{Zeeman}) evaluates to
5$\times$$10^{-21}$, and even 10\% uncertainties in the field value translate  into fractional clock inaccuracies below $10^{-21}$.

{\em Black-body radiation shift.} -- Modern atomic  clocks are susceptible to thermal electromagnetic fields fluctuations. These fluctuations couple to electric and magnetic moments and lead to shifts of clock levels. Usually the dominant contribution comes from the E1 couplings and
then the BBR shift  $\Delta^{(E1)}_{BBR}$ is proportional to the differential electric dipole polarizability and $T^4$, $T$ being ambient temperature~\cite{PorDer06}.
Since the HCI polarizability is reduced as $1/Z_i^4$, the BBR shift is similarly suppressed. Moreover
the differential polarizability for a hyperfine transition is reduced further by the factor of
$\omega_\text{clock}/\omega_{dip}$, where $\omega_{dip} \propto Z_i^2$ is the  frequency of the  E1-allowed transition to the closest level.
This leads to the scaling law  ${\Delta^{(E1)}_{BBR}}/{\omega_\text{clock}}\propto 1/Z_i^{6}$,
resulting in a negligible shift  below $10^{-21}$ for $Z_i>10$  even at the room temperature. While E1 contribution usually overwhelms  the BBR shift, for the M1 clock transitions the dominant contribution comes from the thermal B-field fluctuations that couple to M1 transition moments.
 For a hyperfine transition of Fig.~\ref{Fig1}(a) this M1-BBR shift can be computed as~\cite{Itano}
\begin{equation}\label{BBR_magn}
    \frac{\Delta^{(M1)}_{BBR}}{\omega_\text{clock}}=\frac{16\mu^2_\text{B}}{3\pi\hbar c^3}
\int_{0}^{\infty} \frac{\omega^3 d\omega}{(\omega^2_\text{clock}-\omega^2)(e^{\hbar\omega/k_B T}-1)} .
\end{equation}
For example, for $\omega^{}_\text{clock}/2\pi$=$2\times$$10^{14}$~Hz
at $T=300\, \mathrm{K}$ the resulting relative shift is  5$\times$$10^{-20}$. The fractional clock uncertainty
is below  $10^{-20}$ provided the temperature is stabilized at the  $300 \pm15 \, \mathrm{K}$ level.

{\em AC-Stark shift induced by clock laser.} -- The clock transition has to be ultimately probed by a stable laser that would lock onto the transition.
The probing itself leads to clock shifts due to the AC Stark effect. This effect is proportional to the E1 polarizability discussed in the BBR shift context and
thereby strongly suppressed as  $1/Z^6_i$.

The presented discussion of the optical hyperfine HCI clocks clearly demonstrates that such clock transitions are exceptionally insensitive  to environmental perturbations. As discussed, all common systematic shifts are well below the 10$^{-20}$ fractional accuracy level. Moreover reaching this level of accuracy does not require precision control of ambient conditions. In addition, let us consider the relatively large natural width of the
M1 transitions that imposes some limitations on the choice of candidate ions.
For example, for the hyperfine transition of Fig.~\ref{Fig1}a the natural line-width (in units of Hz) is given by
\begin{equation}\label{gamma}
    \frac{\gamma_{\text{sp}}}{2\pi}=\frac{4\mu^2_\text{B} \omega^{3}_\text{clock}}{6\pi\hbar c^3}.
\end{equation}
Although this quantity may exceed 10 Hz for $\omega_\text{clock}$ in the visible spectrum,
such values of line-widths are hardly a practical limitation in
current optical clock experiments. Indeed, the experiments~\cite{Chou_2010,Huntemann_2012,Madej_2012,Hinkley_2013} used resonances that are probe laser intensity broadened to 1-10~Hz anyway, while the natural line-widths of the clock transitions are several order of magnitudes smaller. In other words, having ultra-narrow clock transitions is not essential -- in fact probing ultra narrow transitions requires ultra-stable lasers with a comparable spectral bandwidth. Because of this practical limitation it is sufficient to focus on optical M1 transitions with natural line-width of 0.1-10 Hz. Eq.(\ref{gamma}) maps this range into  $\lambda_\text{clock}$$=$$2\pi c/\omega_\text{clock}$ in the region between 0.5 to 3 $\mu$m. In particular, for
$\omega_\text{clock}/2\pi$=2$\times 10^{14}$~Hz ($\lambda_\text{clock}$=1.5~$\mu$m) one obtains
$\gamma_\text{sp}$$\approx$1.3~Hz. Longer wavelengths while leading to a desirable reduction of $\gamma_\text{sp}$
also have a negative impact -- they increase fractional inaccuracies and instabilities as  $1/\omega_\text{clock}$.

\begin{table}
\begin{tabular}{|c||c|c|c|c|c|}
\cline{1-3}

\hline
Isotope & $Z$ & $N_e$ & $Z_i$ & $\lambda_\text{clock}$~($\mu$m) & $\gamma_\text{sp}$/2$\pi$~(Hz)\\
\hline
$^{171}$Yb & 70 & 1 & 69 & 2.16  & 0.43 \\
$^{195}$Pt & 78 & 1 & 77 & 1.08  & 3.4 \\
$^{199}$Hg & 80 & 1 & 79 & 1.15  & 2.8 \\
$^{203}$Tl & 81 & 1 & 80 & 0.338 & 111.2 \\
$^{205}$Tl & 81 & 1 & 80 & 0.335 & 114.2 \\
$^{207}$Pb & 82 & 1 & 81 & 0.886 & 6.2 \\
\hline
\end{tabular}
\caption{Hyperfine transitions in hydrogen-like highly-charged ions ($N_e=1$) with
wavelengths  $\lambda_\text{clock}$$<3$~$\mu$m. Here we list stable isotopes with
nuclear spin $I=1/2$. \label{T1}}
\end{table}

In Table~\ref{T1} we list hydrogen-like HCIs that have stable isotopes with nuclear spin 1/2 and
have a hyperfine transition  wavelength  $\lambda_\text{clock}<3\,\mu\mathrm{m}$. These relativistic estimates were carried out using  expressions from Ref.~\cite{Sha94}. Six heavy isotopes satisfy all these conditions: $^{207}$Pb, $^{205}$Tl,
$^{203}$Tl, $^{199}$Hg, $^{195}$Pt, $^{171}$Yb. We find that transitions in Li-like ions of the same isotopes have wavelengths longer than $3\,\mu\mathrm{m}$.
If, however, we consider $I>1/2$ isotopes, the list of candidate HCIs is substantially expanded at the expense of increased degeneracies. The possibility of using the H-like HCIs as qubits was briefly discussed in \cite{Wineland1998}. 

Having discussed the hyperfine clock transitions, now we turn to the {\em fine-structure} manifolds. Below we focus on two illustrative examples (see Figs.~\ref{Fig1}b and \ref{Fig1}c).
All the presented arguments for the hyperfine transitions remain valid for the M1 fine-structure clock transitions, with some minor modifications related to the quadrupolar shifts.

{\em HCIs with a single $p$-valence  electron, Fig.~\ref{Fig1}b.} --
Consider ions with the total number of electrons $N_e$=5,13,31,49 and 63 ground state of which splits into the $^{2}\!P_{1/2}$ and  $^{2}\!P_{3/2}$ fine-structure components.
While the electronic Q-moment of  the  $^{2}\!P_{1/2}$ state vanishes,
 the  $^{2}\!P_{3/2}$  state does have nonzero  Q-moment.
 Nevertheless, the quadrupolar shift can be either eliminated or strongly suppressed by choosing
clock transitions between hyperfine components for isotopes with $I=1$, 3/2 and 2.
For example, for the $I=1$ isotopes we may pick the $|^{2}\!P_{1/2},F=1/2\rangle \to
|^{2}\!P_{3/2},F=1/2\rangle$  transition where the Q-moments simply vanish due to the angular selection rules.  For the $I=2$ isotopes we may choose the transition $|^{2}\!P_{1/2},F=3/2\rangle \to
|^{2}\!P_{3/2},F=1/2\rangle$.  For the $I=3/2$ isotopes the proper choice would be the $|^{2}\!P_{1/2},F=1\rangle\to
|^{2}\!P_{3/2},F=0\rangle$ closed transition (see Fig.~\ref{Fig1}b), where the Q-moment of the state $|^{2}\!P_{3/2},F=0\rangle$ is zero and the Q-moment of the state $|^{2}\!P_{1/2},F=1\rangle$ is very small. Further,  using  the  $|F=1,M_F=0\rangle$  sublevel eliminates the linear Zeeman shift.
Examples of  optical transition for the stable $I=3/2$ isotopes are compiled in the Table~\ref{T2}, while the stable $I=1$ and $I=2$ isotopes do not exist (except for $^{2}$H, $^{6}$Li and $^{14}$N, which are not suitable for our goals).

\begin{table}
\begin{tabular}{|c||c|c|c|c|c|}
\cline{1-3}

\hline
Isotope & $Z$ & $N_{e}$ & $Z_i$ & $^2\!P_{1/2} \to\, ^2\!P_{3/2}$~($\mu$m) & $\gamma_\text{sp}$/2$\pi$~(Hz)\\
\hline
$^{33}$S & 16 & 5 & 11 & 0.761 & 3.25 \\
$^{35,37}$Cl & 17 & 5 & 12 & 0.574 & 7.57 \\
$^{39,41}$K & 19 & 5 & 14 & 0.3446 & 35.0 \\
$^{53}$Cr & 24 & 13 & 11 & 0.8156 & 2.64 \\
$^{61}$Ni & 28 & 13 & 15 & 0.3602 & 30.6 \\
$^{63,65}$Cu & 29 & 13 & 16 & 0.3008 & 52.6 \\
$^{79,81}$Br & 35 & 31 & 4 & 1.642 & 0.32 \\
$^{87}$Rb & 37 & 31 & 6 & 0.9554 & 1.64 \\
$^{131}$Xe & 54 & 49 & 5 & 0.6411 & 5.43 \\
$^{135,137}$Ba & 56 & 49 & 7 & 0.4238 & 18.8 \\
\hline
\end{tabular}
\caption{Fine-structure $np\,^2\!P_{1/2} \to np\,^2\!P_{3/2}$ M1 clock transitions in the optical range 0.3--4~$\mu$m for
$I=3/2$ stable isotopes (wavelength data from \cite{NIST}).}
\label{T2}
\end{table}

{\em HCIs with two $p$ valence-shell  electrons, Fig.~\ref{Fig1}c.} --
The $ (np)^2 \, ^3\!P$ ground state fine-structure manifold of such ions ($N_e$=6,14,32,50,64) have the $^3\!P_0$ lowest-energy state connected to the first excited $^3\!P_1$ state  via an M1 transition.  By choosing isotopes with $I=1/2, 1, 3/2$ one may also either eliminate or substantially suppress the quadrupolar shift. Indeed, for  the $I=1/2$ isotopes the proper choice of the clock transition would be  $|^{3}\!{P}_{0},
F=1/2\rangle \to |^{3}\!{P}_{1}, F=1/2\rangle$ (see Fig.~\ref{Fig1}c) with identically vanishing quadrupolar shift.
For the $I=1$ isotopes one should use the
$|^{3}\!P_{0},$ $F=1$$\rangle \to |^{3}\!P_{1},$
$F=0\rangle$ transition, and for the $I=3/2$ isotopes -- the $|^{3}\!P_{0},$ $F=3/2$$\rangle \to
|^{3}\!{P}_{1},$ $F=1/2$$\rangle$ transition. Compared to the single $p$-electron case of Fig.~\ref{Fig1}b, an added benefit of  such HCIs is the simple single-component structure of the ground state which simplifies the initial state preparation. Suitable HCIs for $I=1/2$ and 3/2 stable isotopes are listed in  Tables~\ref{T3} and~\ref{T4}, respectively.
The linear Zeeman shift can be eliminated by averaging over two clock transitions with opposite magnetic quantum numbers $M_F$. In particular, in the $I=1/2$  case we can use  a set of two transitions $|^{3}\!P_{0},F=1/2,M_F=\pm 1/2\rangle \to |^{3}\!{P}_{1},F=1/2,M_F=\pm 1/2\rangle$   probed with linearly polarized  laser.

\begin{table}
\begin{tabular}{|c||c|c|c|c|c|}
\cline{1-3}

\hline
Isotope & $Z$ & $N_{e}$ & $Z_i$ & $^3\!P_0 \to\, ^3\!P_1$~($\mu$m) & $\gamma_\text{sp}$/2$\pi$~(Hz)\\
\hline
$^{29}$Si & 14 & 6 & 8 & 3.929$^{(a)}$ & 0.05 \\
$^{31}$P & 15 & 6 & 9 & 2.709$^{(a)}$ & 0.14 \\
$^{57}$Fe & 26 & 14 & 12 & 1.075$^{(a)}$ & 2.3 \\
$^{89}$Y & 39 & 32 & 7 & 1.120$^{(b)}$ & 2.04 \\
$^{103}$Rh & 45 & 32 & 13 & 0.321$^{(b)}$ & 86.5 \\
$^{123,125}$Te & 52 & 50 & 2 & 2.105$^{(a)}$ & 0.31 \\
$^{129}$Xe & 54 & 50 & 4 & 1.076$^{(a)}$ & 2.3 \\
\hline
\end{tabular}
\caption{Fine-structure $np^2\,^3\!P_0 \to np^2\,^3\!P_1$ M1 clock transitions in the optical range 0.3--4~$\mu$m for
$I=1/2$ stable isotopes ($^{(a)}$ -- data from \cite{NIST}, $^{(b)}$ -- data from \cite{BieHimGar1990}).}
\label{T3}
\end{table}

\begin{table}
\begin{tabular}{|c||c|c|c|c|c|}
\cline{1-3}

\hline
Isotope & $Z$ & $N_{e}$ & $Z_i$ & $^3\!P_0 \to\, ^3\!P_1$~($\mu$m) & $\gamma_\text{sp}$/2$\pi$~(Hz)\\
\hline
$^{33}$S & 16 & 6 & 10 & 1.9201 & 0.4 \\
$^{35,37}$Cl & 17 & 6 & 11 & 1.381 & 1.1 \\
$^{39,41}$K & 19 & 6 & 13 & 0.75557 & 6.64 \\
$^{53}$Cr & 24 & 14 & 10 & 1.806 & 0.49 \\
$^{61}$Ni & 28 & 14 & 14 & 0.67035 & 9.5 \\
$^{63,65}$Cu & 29 & 14 & 15 & 0.5377 & 18.4 \\
$^{69,71}$Ga & 31 & 14 & 17 & 0.35599 & 63.4 \\
$^{79,81}$Br & 35 & 32 & 3 & 3.8138 & 0.05 \\
$^{87}$Rb & 37 & 32 & 5 & 1.9455 & 0.39 \\
$^{131}$Xe & 54 & 50 & 4 & 1.0762 & 2.3 \\
$^{135,137}$Ba & 56 & 50 & 6 & 0.64488 & 10.7 \\
\hline
\end{tabular}
\caption{Fine-structure $np^2\,^3\!P_0 \to np^2\,^3\!P_1$ M1 clock transitions in the optical range 0.3--4~$\mu$m for
$I=3/2$ stable isotopes (wavelength data from \cite{NIST}).}
\label{T4}
\end{table}

It is worth noting that only the clock transitions are in the optical spectral region
and in HCIs all the E1-allowed transitions are shifted away into the ultra-violet or x-ray regions thus making traditional laser cooling difficult. An alternative is to use sympathetic cooling
with co-trapped low-charge ions which would be laser cooled on their E1-allowed transitions.
For example, Ref.~\cite{Gruber_2005} experimentally demonstrated
sympathetic cooling of Xe$^{44+}$ with Be$^{+}$ ions.  In terms of the charge-to-mass ratio $Z_i/M_i$,
$M_i$ being the HCI mass choosing HCIs with  fine-structure clock transitions may be preferable over hyperfine clock transitions as the latter require higher degree of ionization.
One could also use the co-trapped ion
for quantum-logic spectroscopy~\cite{schmidt05}  for registering clock transitions in HCIs.
If the natural line width is larger than 100 Hz, one could employ the continuous wave spectroscopy by
detecting spontaneously emitted photons. In that case one could simultaneously continue  the process of laser cooling of co-trapped low-charge ion as the AC-Stark shifts of the clock transition are negligible.
The time dilation effect \cite{Chou_2010_2} of special relativity
for the HCI clocks was discussed in \cite{Derevianko_2012}. As an additional advantage of using M1 transition we point out the applicability of direct frequency comb spectroscopy
\cite{Fortier_2006}, as for these transitions the required intensity of the probe field is low
($\ll \mu$W/cm$^2$ for Rabi frequencies at the level of 100~Hz and below).

Finally, we would like to comment on sensitivity to the variation of fundamental constants.
 One could parameterize the variation of the clock frequency as \cite{FlaDzu09} ${\delta(\omega_\mathrm{clock}/U)}/{ (\omega_\mathrm{clock}//U)} = \delta V/V$ with
\begin{equation}\label{Eq:DetailedCoeff}
 V =  \alpha^{K_\alpha} \left( \frac{ m_q}{\Lambda_\mathrm{QCD}} \right)^{K_q}  \left(\frac{m_e}{m_p}\right)^{K_{m_e/m_p}} \, .
\end{equation}
Here $U$ is the unit of angular frequency, $m_q$ is a quark mass, $\Lambda_\mathrm{QCD}$ is the quantum chromodynamics mass scale, and $m_e/m_p$ is the electron to proton mass ratio.
For hyperfine transitions the sensitivity coefficients  $K_{m_e/m_p}=1$ with $K_q$ and  $K_\alpha$  depending on the nuclear and atomic structure.
Moving to HCIs has shifted hyperfine transitions from the microwave to the optical spectrum and also made the valence electrons more susceptible to strong relativistic effects. To quantify this effect
 we  evaluated $K_\alpha$ analytically  for hyperfine transitions in the ground state of H-like ions
\begin{eqnarray}\label{Ka}
K_\alpha=\frac{ 6+ (\alpha Z)^2 \left(2 \sqrt{1- (\alpha  Z)^2}-7\right)}{3 -7 (\alpha Z)^2+ 4 (\alpha  Z)^4} \,.
\end{eqnarray}
This fully relativistic estimate does not include QED or finite nuclear size effects.
For low-charged ions $K_\alpha \approx 2$, however $K_\alpha$ rapidly grows for HCIs, reaching infinite values for $Z=\sqrt{3}/(2\alpha)\approx 118.7$. For H-like $^{207}\mathrm{Pb}^{81+}$, $K_{\alpha}=4.04$.
Notice that  usually-employed hyperfine clock transitions lie in the microwave spectral region. As our proposal moves such transitions to the optical domain it expands the range of experimental schemes for probing  space and time variations  of fundamental constants.

To summarize, compared to the proposals of using high-multipole transitions for the HCI clockwork~\cite{Derevianko_2012,Safronova_2014}},
we argue that the most direct advantage of our proposal comes from the low degeneracy of clock levels and the simplicity of atomic structure in combination with negligible quadrupolar shift. This simplicity is anticipated to translate into easier experimental clock initialization, probing, and resonance detection.
In addition to such ``Occam's razor'' appeal of the M1 HCI clockwork, there is a substantial natural suppression or elimination of quadrupole shifts for clock transitions of Fig.~\ref{Fig1}. This fact can be a key advantage for the next generation of ion clock, because it again simplifies experimental realization and does not need to rely on applications of special averaging techniques~\cite{Ita00,DzuDerFla12HICClockSpinZero}.

{\em Acknowledgements --} We thank V.M.~Shabaev for useful discussions.  V.I.Yu. and A.V.T. were supported by the RFBR (grants 14-02-00712, 14-02-00939, 14-02-00806), by the Russian Academy of Sciences and
Presidium of the Siberian Branch of the Russian Academy of
Sciences. A.D. was supported by the US National Science Foundation.

\end{document}